\documentclass[11pt,twoside]{article}
\usepackage{asp2006}
\usepackage{psfig}
\usepackage{epsf}
\usepackage{graphics}
\usepackage{lscape}
\def\edcomment#1{\iffalse\marginpar{\raggedright\sl#1\/}\else\relax\fi}
\reversemarginpar

\begin{document}
\title {Co-evolution of AGN and Star-forming Galaxies in the Australia Telescope Large Area Survey
}

\author{Ray P. Norris}
\affil{CSIRO ATNF, PO Box 76, Epping, NSW 1710, Australia}

\begin{abstract} ATLAS (Australia Telescope Large Area Survey) is a wide deep  radio survey which is distinguished by its comprehensive multi-wavelength approach. ATLAS is creating a large dataset of radio-selected galaxies for studying the evolution and inter-relationship of star-forming and active galaxies. Although the project is far from complete, we are already starting to answer some of these questions, and have stumbled across three surprises along the way: 
\begin{itemize}
\item{FRI/FRII radio-loud AGN embedded within spiral galaxies, }
\item{radio-bright AGN which are unexpectedly faint in the infrared, and which may be at high redshift}
\item{IR-luminous radio-quiet AGN which are partly responsible for the wide variations in reported values of the radio-infrared ratio}
\end{itemize}
These and other observations suggest that the AGN activity and star formation become increasingly inter-dependent at high redshifts.
\end{abstract}

\section{Introduction}
Multiwavelength deep wide surveys are a key tool for studying galaxy evolution. Radio surveys, once a niche element of these studies, are becoming increasingly important because of their increasing sensitivity, because they access large fields of view, because they are unaffected by dust, and because they can detect active galactic nuclei (AGN) up to the highest redshifts. The Australia Telescope Large Area Survey (ATLAS) is surveying a seven square degree area with a target rms of 10-15 $\mu$Jy rms. The main science goals of ATLAS are to help determine how galaxies evolved through cosmic time, and help determine the relationship between star formation and massive black holes.

The power of deep radio surveys can only be realised by corresponding data at other wavelengths. The area surveyed by ATLAS has been chosen to overlap that observed by the Spitzer Wide-area Infrared Extragalactic Survey (SWIRE) program (Lonsdale et al. 2003), around the CDFS and ELAIS-S1 regions, and include the GOODS and ECDFS regions. As a result, the ATLAS radio data are accompanied by extensive infrared, and optical data. Furthermore, the ATLAS fields will also be observed in the optical and infrared by future deep survey projects such as VISTA/VIDEO (Jarvis 2008), Spitzer/SERVS (Lacy et al. 2008) and Herschel/Hermes (S. Oliver, private communication).

In addition to these astrophysical goals, ATLAS is an important pathfinder for the much more extensive studies that will be possible with next--generation telescopes such as ASKAP (Johnston et al., 2008). The proposed ASKAP/EMU project, which aims to survey 75\% of the sky to a depth 40 times greater than that of NVSS (Condon et al. 1998) is a direct descendent of ATLAS. 

\section{Current Status}

We are now part way through the ATLAS observations, having covered 7 square degrees of the CDFS and ELAIS-S1 SWIRE fields to an rms sensitivity of about 20-30 $\mu$Jy, and a spatial resolution of about 10 arcsec. From these images we have identified and catalogued about 2000 radio sources, and have made the data products publicly available (Norris et al. 2006, Middelberg et al. 2008a). 

In addition to these total intensity images, we are also measuring polarisation of all sources (Hales et al., in preparation), spectral index between 20 and 13 cm (Middelberg et al., in preparation), and have started an extensive VLBI program.

Nearly all the radio sources have been detected by Spitzer, and most have optical photometry, giving up to 10 bands of photometry for Spectral Energy Distribution (SED) fits and photometric redshifts. 

\begin{figure}[]
\begin{center}
\epsfysize=70mm
\epsfbox[10 10 1000 300]{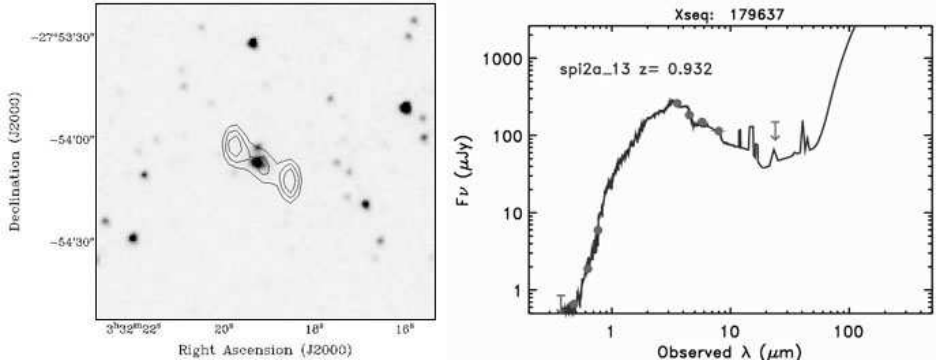}
\caption{The composite source S425. The left image shows radio contours superimposed on a 3.6 $\mu$m SWIRE image, and the right image (Mari Poletta, private communication) shows the SED. This source has the radio morphology of an AGN, but the SED of a star-forming galaxy.}
\end{center}
\end{figure}

Although the rms noise in the ATLAS images is largely close to the theoretical limit, the published data are degraded in some places by imaging artefacts which are resistant to standard tools such as cleaning, selfcal, and peeling. Lenc et al (in preparation) have recently developed techniques to remove these artefacts, and we plan a data release in 2009 which will be free of these artefacts.  

To study the evolution of galaxies obviously requires us to know their distances, and so ATLAS includes a major spectroscopic program (Mao et al., 2009). So far we have spectroscopic redshifts for about 25\% of the ATLAS sources, with redshifts as high as z$\sim$3. We are also studying particular subsets of sources such as Gigahertz Peaked Spectrum (GPS) sources (Randall et al. 2009) and head-tail galaxies (Mao et al. in preparation), as diagnostic probes of galaxy evolution and structure. Within this extensive program, we have so far found three surprising results.

\section{First surprise: star-formation or AGN?}

\begin{figure}[]
\begin{center}
\epsfysize=60mm
\epsfbox[0 10 1000 400]{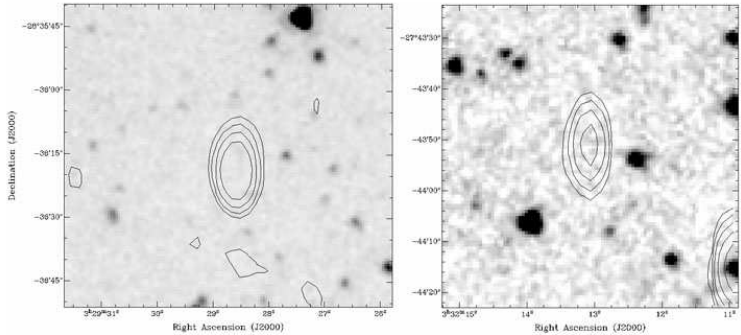}
\caption{Two Infrared-Faint Radio Sources, taken from Norris
et al. (2006). Contours show 20 cm radio emission, and greyscale shows 3.6
$\mu$m infrared emission.}
\end{center}
\end{figure}

Attempts to understand the evolution of galaxies continue to be plagued by the problem of distinguishing between star-formation activity and AGN. While optical emission line ratios remain the standard discriminant for low-redshift galaxies, they are unreliable for dusty galaxies, which constitute a substantial fraction of galaxies in deep radio surveys. A key goal of the project is to test a variety of potential discriminants to distinguish between AGN and star-forming activity (Randall et al., 2009).

At low redshifts, high-powered radio galaxies ($> 10^{25} W Hz^{-1}$) are almost always associated with elliptical galaxies. In the ATLAS sample, we have so far found about 50 candidate sources with the SED of a star-forming galaxy, but the radio power and morphology of a powerful radio galaxy.
For example, Fig. 1 shows source S425, at a photometric redshift of 0.932, which has a 20cm flux density of 9 mJy, corresponding to a luminosity of luminosity of $4 \times 10^{25}$ WHz$^{-1}$, placing it close to the FRI/FRII break, and a classical triple radio morphology. However, its SED and optical morphology resemble a star-forming spiral galaxy. Such objects are rare in the local Universe, with only two nearby examples: 0313-192 (Keel et al. 2006) and NGC612 (Ekers et al. 1978). 

The ATLAS catalogue includes several examples of galaxies whose SED is of a star-forming spiral galaxy, but which have the radio luminosity or morphology of an AGN. We suggest that these represent a class of AGN buried deeply inside a dusty star-forming galaxy. This phenomenon appears to be increasingly common at high redshifts.

\section{Surprise Two: Infrared-Faint Radio Sources (IFRS)}

\begin{figure}[]
\begin{center}
\epsfysize=100mm
\epsfbox[0 10 1000 1000]{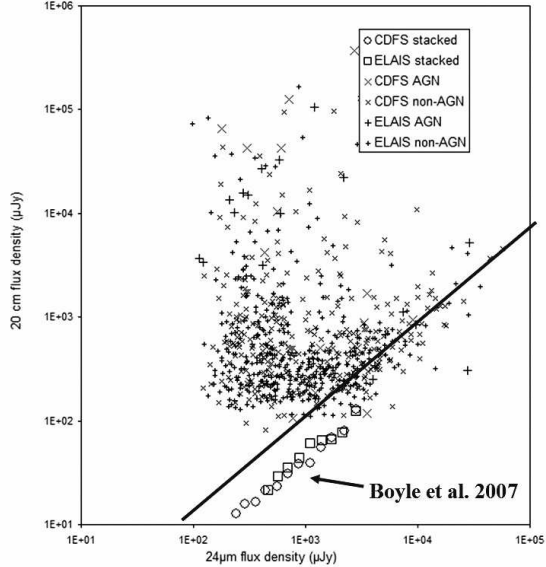}
\caption{The radio-infrared correlation, adapted from  Norris
et al. (2008). The diagonal line shows the "standard" radio-IR correlation, and the squares represent the stacked value from Boyle et al. 2007.}
\end{center}
\end{figure}

Unexpectedly, we have found 53 radio sources (IFRS: Infrared-Faint Radio Sources) as strong as 20 mJy at 20 cm (Norris et al. (2006) , Middelberg et al. (2008a)), which are have
no observable infrared counterpart in the SWIRE survey. Two examples are shown in Fig. 2. These IFRS were unexpected as SWIRE was
thought to be deep enough to detect all ATLAS radio sources
in the local universe, regardless of whether star formation
or AGN powered the radio emission.
Norris et al (2006) have shown that IFRS are undetected in stacked Spitzer images, so that if they represent a tail of the standard infrared flux density distribution, the distribution is either bimodal, or else has a very long tail.

Two of these sources (Norris et al. 2007; Middelberg et al. 2008b) have been detected using VLBI, implying that they have AGN cores.  Further evidence for an AGN origin comes from the wide range (-2 to +1) of spectral indices. 

Garn \& Alexander (2008) stacked IFRS sources
in the Spitzer First Look Survey (FLS: Rieke et al., 2004) at infrared wavelengths
as well as at 610 MHz. They find that the IFRS sources
can be modelled as compact Fanaroff Riley type II (FRII)
radio galaxies at high redshift (z $\sim$ 4).

Four IFRS lie within the extended CDFS for which
there is now deeper infrared data available. Huynh et al. (in preparation) have detected two IFRS
sources in these more sensitive IRAC images. The non-detection of two other IFRSs strongly constrains the source type, and SED modelling shows that these objects appear to be high redshift (z $>$ 2).
radio-loud AGN

\section{Surprise Three: The radio-FIR correlation}

The well-known  radio-far-infrared correlation (e.g. de Jong et al. 1985) which is seen for low-redshift star-forming galaxies is reflected in a similar correlation between the 24 $\mu$m and 20 cm emission. Appleton et al. (2004) have used data from the FLS to show this correlation extends out to z $\sim$ 1, and Beswick et al. (2008) have shown that the luminosities of all identified 24 $\mu$m sources still follow the correlation down to S$_{24{\mu}m} \sim 80 \mu$Jy. These observations have been interpreted as a function of evolution by Seymour et al. (2008). To explore the correlation at even fainter levels, Boyle et al. (2007) stacked ATLAS radio data to show that it extends down to microJy levels.

Fig. 3 shows all ATLAS sources with Spitzer 24 $\mu$m flux densities. They include both star-forming galaxies, which roughly follow the correlation found by Appleton et al., and AGN, which extend to the top of the plot. Also shown are the stacked data from Boyle et al. Although the stacked data broadly follow the correlation, they follow a value of q=log(S $_{24\mu m}$/S$_{20cm}$)=1.4 which is significantly different from the q=0.8 found by Appleton et al. Further exploration has found that this same result can be reproduced by stacking data from the FLS (Norris et al. in preparation). The difference between the Boyle et al result and the other results is not caused by an error in the data or in the stacking algorithm, but by the selection of the sources. Infrared-selected sources give a value of q similar to that obtained by Boyle et al., whilst radio-selected data give a value of q similar to that obtained by Appleton et al. Norris et al. have shown that the difference is at least partly attributable to a population of infrared-excess sources which have a high value of q. They appear on optical images as galaxies or quasars, and spectroscopy has shown their redshifts to extend to above 3. They appear to be a hitherto-unrecognised population of radio-quiet AGN, in which the excess IR emission may be caused by a heated dust torus.

\section{Conclusion and Future Work}

Because of their ability to penetrate dust and reveal AGNs, and their increasing sensitivity and area, deep wide radio surveys like ATLAS are starting to have a profound impact on our understanding of galaxy evolution, and are particularly important for finding populations of rare but important objects.

ATLAS has so far found three such classes of objects which were previously unrecognised:
\begin{itemize}
\item{Radio-loud AGNs embedded in star-forming galaxies}
\item{Infrared-Faint Radio Sources}
\item{A largely unrecognised population of IR-excess radio-quiet AGN  which are partly responsible for the differences in reported values of the radio-infrared correlation coefficient}
 \end{itemize}

\subsection*{Acknowledgements}

I thank the entire ATLAS team, listed on \\
http://www.atnf.csiro.au/research/deep/team/index.htm, for their contributions to the work summarised here.

\subsection*{References}
Appleton, P. N., et al. 2004, ApJS, 154, 147\\
Beswick, R. J., et al.,\ 2008, \mnras, 385, 1143 \\
Boyle, B., et al., 2007, MNRAS, 376, 1182B\\
Condon, J. J., et al. 1998, AJ, 115, 1693\\
de Jong T., Klein,U., Wielebinski,R., Wunderlich,E.,1985, AA, 147, 6\\
Ekers R. D., Goss W. M., Kotanyi C. G., \& Skellern D. J., 1978, A\&A, 69, L21\\
Garn, T., \& Alexander, P.\ 2008, \mnras, 391, 1000 \\
Jarvis, M.~J.\ 2008, arXiv:0801.4618 \\
Johnston, S., et al.\ 2008, Experimental Astronomy, 22, 151\\
Keel, W.C., et al. 2006, AJ, 132, 2233\\
Lacy, M., et al.\ 2008, , Spitzer Proposal ID \#60024\\
Lonsdale, C. J., et al.\ 2003, PASP, 115, 897\\
Mao, M.Y., et al.\ 2009, ASPC, these proceedings\\
Middelberg, E., et al., 2008a, \aj, 135, 1276\\
Middelberg, E., et al., 2008b, \aap, 491, 435 \\
Norris, R.~P., et al.\  2006, AJ, 132, 2409\\
Norris R. P., et al.\ 2007, MNRAS, 378, 1434\\
Norris R. P., et al.\ 2008, astro-ph/0804.3998\\
Randall, K.E., et al.\ 2009, ASPC, these proceedings\\
Rieke G. H., et al., 2004, ApJS, 154, 25\\
Seymour, N., et al.\ 2008, \mnras, 386, 1695

\end{document}